\documentclass{llncs}
\usepackage[margin=1in]{geometry}

\usepackage[inline]{enumitem}

\usepackage{subcaption}
\usepackage{booktabs}
\usepackage[hyphens]{url}
\usepackage{hyperref}

\usepackage{xcolor}
\definecolor{labelcolor}{cmyk}{0.22,0.10,0.10,0.10}
\definecolor{listbackgroundcolor}{cmyk}{0.10,0.10,0.05,0.05}
\definecolor{listbackgroundcolorlight}{rgb}{0.91,0.92,0.94}

\usepackage{listings}
\usepackage{tikz}
\usetikzlibrary{shapes}
\usetikzlibrary{calc}
\usetikzlibrary{arrows.meta}

\tikzstyle{message}=[->,-{Classical TikZ Rightarrow[length=1.5mm]}]

\usepackage{xspace}
\newcommand{\etal}{{et al.\@\xspace}}

\usepackage{amssymb}
\newcommand{\ul}{\ulcorner}
\newcommand{\ur}{\urcorner}
\newcommand{\inn}{\ensuremath{\ul\mathsf{in}\ur}\xspace}
\newcommand{\out}{\ensuremath{\ul\mathsf{out}\ur}\xspace}
\newcommand{\nil}{\ensuremath{\ul\mathsf{nil}\ur}\xspace}
\newcommand{\mo}{\mapsto}

\newcounter{mpscount}
\newcounter{akccount}
\newcounter{shccount}

\newcommand{\fbf}{\textbf}
\newcommand{\fsl}{\textsl}
\newcommand{\fsf}[1]{{\footnotesize{\textsf{#1}}}}

\newcommand{\mname}[1]{\fsl{#1}}
\newcommand{\pname}[1]{\fsl{#1}}
\newcommand{\rname}[1]{\textsc{#1}}
\newcommand{\val}[1]{\texttt{#1}}
\newcommand{\paraname}[1]{\fsf{#1}}

\DeclareRobustCommand{\nUmErAL}[1]{#1}

\begin{document}

\title{Toolsuite for Implementing Multiagent Systems Based on Communication Protocols}

\author{Amit K.~Chopra\orcidID{0000-0003-4629-7594},\\
Samuel H. Christie V\orcidID{0000-0003-1341-0087}, and\\
Munindar P.~Singh\orcidID{0000-0003-3599-3893}}

\institute{Lancaster University, UK,\\
\email{amit.chopra@lancaster.ac.uk}\\ 
North Carolina State University,\\
\email{schrist@ncsu.edu}\\
North Carolina State University,\\
\email{singh@ncsu.edu}}

\maketitle

\abstract{Interaction-Oriented Programming (IOP) is an approach to building a multiagent system by modeling the interactions between its roles via a flexible interaction protocol and implementing agents to realize the interactions of the roles they play in the protocol.
\hfill\break\hspace*{\parindent}
In recent years, we have developed an extensive suite of software that enables multiagent system developers to apply IOP. These include tools for efficiently verifying protocols for properties such as liveness and safety and middleware that simplifies the implementation of agents. This paper presents some of that software suite.
}

% 1) A clear motivation for the need of the Agent Toolkit

% 2) A "semi-systematic" literature review covering the literature on tools that share similar features with the Agent Toolkit (at least 15-20 papers also from known sources in the community: AAMAS, JAAMAS, EMAS, EUMAS)

% 3) A technical description of the Agent Toolkit, at a level of abstraction that allows even non expert readers to grasp its novelty, usefulness, and potential for application.

% 4) A critical discussion of limitations and technical challenges, also in comparison with related solutions.

% 5) A critical discussion of how the Agent Toolkit positions w.r.t.:
% 5.1) Good Old Fashioned Autonomous Agents Engineering, and Good Old Fashioned Engineering multiagent systems
% 5.2) Generative AI / Generative Agents
% 5.3) Ethical issues

% 6) A Blue-sky vision

\section{Motivation}

Software systems increasingly support interactions among autonomous principals (humans and organizations) in various settings, e.g., e-commerce, healthcare, finance, and so on.
Owing to autonomy, it is natural to realize such systems as multiagent systems in which principals are represented by agents who interact by exchanging messages.

The main challenges in realizing multiagent systems can be divided into broad categories.
One, how can we model a multiagent system in a manner that enables realizing it as a loosely-coupled system of agents who can interact with each other with maximal flexibility.
Intelligent actions by agents presumes flexibility; it means an agent can take into account the relevant circumstances in making decisions.
Loose coupling \cite{parnas:design:1971,Shaw+Garlan-96} means minimizing and making explicit the architectural assumptions that agents (as system components) make of each other.
Loose coupling thus promotes architectural simplicity, clarity, and efficiency.
Crucially, it enables implementing an agent without knowing how other agents are implemented (which in open systems may not be possible anyway).

Two, given a model of a multiagent system, we are faced with the problem of implementing an agent to play a role in it.
How can we exploit the model to facilitate such engineering? In particular, how can we engineer flexible, fault-tolerant agents that are correct with respect to the model in a manner that enables programmers to focus on encoding an agent's decision making.

Modeling multiagent systems, first and foremost, in terms of \emph{interaction protocols} is the most promising way we know of addressing these challenges.
This approach offers several benefits.
First, a protocol supports autonomy by capturing communication constraints but otherwise leaving the principals free to apply their own business logic and engage flexibly.
Second, a protocol helps principals implement their software agents by providing role-based interfaces.
Third, protocols enable realizing software in terms of loosely coupled, decentralized components (the agents).
Fourth, protocols may be composed and verified, which enables reasoning about a system before any agents are implemented \cite{ProMAS-invited-09}.
Fifth, they support high-level abstractions such as social commitments \cite{singh:rethinking:1998} and, more generally, norms, which inform agent decision making.

The importance of interaction protocols was recognized early on in multiagent systems research \cite{Gasser-91,Hewitt91}, which spurred research on languages for specifying protocols.
AUML \cite{Odell+2001} was a notable early result of this activity and was highly influential, finding application in communication standards and software methodologies.
However, AUML and most of the other work (reviewed in Section~\ref{sec:literature}) that followed it did not enable realizing the above-mentioned benefits.

The situation changed with the development of information protocols \cite{AAMAS-BSPL-11,ICWS-LoST-11,AAMAS-BSPL-12,SCC-14:Bliss}, a novel declarative approach for specifying multiagent protocols.
Traditionally, protocols specified message ordering.
Information protocols depart from this tradition by specifying information causality and integrity constraints on communication.
In recent years, we have built a software suite that enables verifying information protocols and implementing flexible, robust agents based on high-level information-based abstractions.

Our goal in this contribution is to show via concrete examples how this suite enables realizing all of the above-mentioned benefits.

\section{Literature}
\label{sec:literature}

We review the literature on related themes.

\subsection{Agent Communication Languages}
Agent communication is usually understood as building on elements of the philosophy of language, especially \emph{speech act theory} \cite{Austin62}.
Here, a communication is an action performed by the speaker (depending on context).
For example, if a referee in a soccer match says ``foul,'' it's a foul (and changes the state of the game), but if a player says ``foul,'' it has no such effect.
If a fan skeets ``foul by Smith'' they may inform (or misinform) their readers, but cannot cause a foul to be recorded.

Speech act theory separates propositional \emph{content} (e.g., ``foul'') from \emph{illocutionary type} (e.g., \fsl{declarative} to make the content true or \fsl{informative} to report on the content).
Traditional AI approaches define a handful of message formats, one for each major illocutionary type, which is inherently limited since there are potentially as many sets of illocutions as multiagent systems \cite[Section~7]{AC-Manifesto-13}.
KQML (Knowledge Query and Manipulation Language) \cite{Finin+94} was designed for agents viewed as homogeneous knowledge bases (KBs).
The agents can query each other's KBs and tell each other facts to be believed, as well as give commands to be achieved.
FIPA  \cite{Poslad-07:FIPA} is a successor to KQML: streamlined but similar in spirit to it \cite{singh:rethinking:1998}.

Modeling meaning is essential for reasoning about interactions.
For example, a ``quote'' may refer to 
\begin{enumerate}
\item the last trading price on a stock exchange
\item an offer to sell
\end{enumerate}
Yet, current approaches specify meaning only informally.
The early formal models, such as FIPA and KQML, follow a \emph{mentalist} semantics \cite{Searle69}, interpreting communicative acts in terms of the beliefs and intentions of the participants.
Such models are unsuited to autonomous and heterogeneous agents, whose internal representations are hidden.

We established the contrary \emph{social} semantics in AI \cite{singh:rethinking:1998}, about the social semantics of a communicative act based on the commitments it presupposes and alters \cite{Ijcai-99-ACL}.
Modern philosophy of language is reviving elements of Austin's theory pertaining to conventions and norms \cite{Caponetto+Labinaz-23:Sbisa-festschrift,Sbisa-07,Sbisa-18:speech-norms}, which were downplayed in the purely cognitive approaches.
When the social state is expressed via norms, communications map to changes in norms and can be formally reasoned about \cite{AAMAS-16:Custard}.
For example, an offer by a seller creates a commitment; a buyer's acceptance of an offer makes the buyer committed to paying the offered price and advances the seller's commitment to one where the seller has to supply the goods.

% Commitments help formalize communicative acts \citep{Fornara+07} and provide an alternative basis for FIPA messages \citep{Fornara+Colombetti-03}.

% \citet{Rovatsos+15:protocols} propose a formal operational semantics for communications based on commitments under synchronous messaging.

% Their approach violates autonomy by legislating agent behaviors from within the communication protocol, a level of prescription ill-suited to multiagent applications.

% Argumentation \citep{Norman+04} involves maintaining a commitment store---usually centralized, accessed via synchronous messaging, and semiformally specified.

% Other researchers have built on our semantics, e.g.,
% \citep{Winikoff+Liu+Harland-05, Bentahar+04, Bentahar+07, AC-Manifesto-13, Baldoni+18:type-checking, Kafali+Torroni-18:delegation}.

\subsection{Protocol Languages}

Protocols are crucial to multiagent systems engineering methodologies \cite{Cernuzzi+Zambonelli-04:Gaia,AOIS-05,Winikoff+Padgham-05,Rooney+04:VIPER}.
However, protocols are traditionally expressed in informal UML-inspired notations such as \emph{Agent UML} \cite{Huget+Odell-04:AUML}.
FIPA \cite{FIPA-IP} specifies a select few protocols using AUML.
An agent plays a role in a protocol and communicates accordingly.

However, FIPA and UML provide no formal model of the protocol.
Since they lack a formal semantics, it is not possible to verify protocols for desirable properties, provide a principled programming model, or check whether each message respects the protocol semantics.
Moreover, the few protocols that FIPA specifies cannot meet the requirements of the potentially infinite variety of multiagent systems.
Thus, they fall short of the goals for engineering multiagent systems \cite{Winikoff-commitment-07}.

Formal protocol specification approaches generally express the information content implicitly and coordination explicitly, which limits flexibility.
Baldoni {\etal} \cite{Baldoni+06,AAMAS-Choice-09} specify protocols as state machines where the transitions represent messages.
Ferrando {\etal} \cite{Ferrando+19:enactability} specify protocols as trace expressions over messages.
Winikoff {\etal} \cite{Winikoff+18:HAPN} specify protocols via a notation reminiscent of statecharts and augmented with information constraints.
ASEME \cite{Spanoudakis+Moraitis-22:ASEME} is another model inspired by statecharts.
Although these approaches can be applied toward engineering nominally decentralized multiagent systems, these approaches are neither conducive to loose coupling nor flexibility.

\subsection{Agent Programming}

JADE \cite{Bellifemine+07:JADE,Bergenti+20:JADE}, a programming model for multiagent systems, is noteworthy for its early support for FIPA protocols \cite{FIPA-IP}; however, as discussed above, the FIPA approach is long outdated \cite{singh:rethinking:1998} and the FIPA protocols are limited to a few patterns of interaction specified in terms of message ordering.
SARL \cite{Galland+20:SARL} is an imperative language for agents; it lacks support for protocols.

Agent-oriented programming models such as Jason \cite{Bordini+Huber-10:Jason} and JaCaMo \cite{Boissier+13:JaCaMo} provide cognitive abstractions for encoding an agent's internal reasoning but do not support protocols.
Jason uses KQML-inspired communication abstractions; JaCaMo includes Jason and, in addition, supports communication between agents via Web services-style artifacts.

\section{Toolkit Description}

We first introduce the idea of information protocols.
Then we introduce related tooling.
First, we describe a tool for verifying information protocols.
Then, we describe protocol-based programming models, specifically Kiko \cite{AAMAS-23:Kiko} and Mandrake \cite{JAAMAS-22:Mandrake}.
Kiko demonstrates how a generic information-based adapter abstracts over the network and presents a simple information-based interface for programming agents.
Mandrake shows how agent developers may specify application-level message forwarding policies to deal with potentially lost messages.

\subsection{Information Protocols}

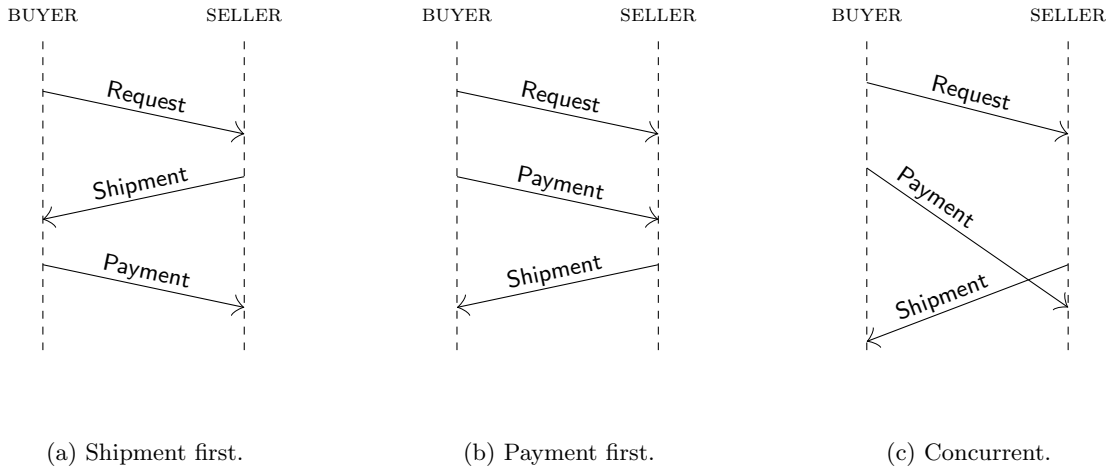
\begin{figure}[htb!]
 \tikzstyle{role}=[thin,draw,align=center,font=\small\sffamily,rectangle,anchor=center,minimum height=5ex,minimum width=6ex,inner sep=1]

 \tikzstyle{every text node part/.style}=[align=center]

 \tikzset{every node text/.style={node contents=\transform{#1}}}
\newcommand{\transform}[1]{\ensuremath{\mathsf{#1}}}

 \tikzstyle{mLabelbase}=[draw=none,midway,fill=none,sloped,align=center,font=\small\sffamily]
 \tikzstyle{mLabelup}=[mLabelbase,above=-2pt]
 \tikzstyle{mLabeldown}=[mLabelbase,below=-2pt]

 \tikzstyle{a_label}=[draw=none,sloped,fill=white,align=center,font=\small\scshape]

% \tikzstyle{message}=[->, >=stealth']
 \tikzstyle{emptybox}=[draw=none,minimum height=3ex]

\centering
\begin{subfigure}[t]{0.33\linewidth}
\centering
\begin{tikzpicture}
\matrix () [row sep=10,column sep=70] {
\node[emptybox] (a) {};
& \node[emptybox] (b) {};\\
\node (a-zero) {};
& \node (b-zero) {}; \\
\node (a-one) {};
& \node (b-one) {}; \\
\node (a-two) {};
& \node (b-two) {}; \\
\node (a-three) {};
& \node (b-three) {}; \\[1]
\node (a-four) {};
& \node (b-four) {}; \\
\node (a-five) {};
& \node (b-five) {}; \\
\node (a-six) {};
& \node (b-six) {}; \\
\node[emptybox] (ae) {};
& \node[emptybox] (be) {};\\
};
\node [role,draw=none,anchor=south] at (a) {\rname{buyer}};
\node [role,draw=none,anchor=south] at (b) {\rname{seller}};

\draw [dashed] (a.center)--(a-six.center);
\draw [dashed] (b.center)--(b-six.center);

\draw [message] (a-zero.center)--node [mLabelup] {Request} (b-one.center);

\draw [message] (b-two.center)--node [mLabelup] {Shipment} (a-three.center);

\draw [message] (a-four.center)--node [mLabelup] {Payment} (b-five.center);

\end{tikzpicture}
\caption{Shipment first.}
\label{fig:shipment-first}
\end{subfigure}% <== Mandatory %
\hfill
\begin{subfigure}[t]{0.33\linewidth}
 \centering
\begin{tikzpicture}
  \matrix () [row sep=10,column sep=70] {
 \node[emptybox] (a) {};
 & \node[emptybox] (b) {};\\
 \node (a-zero) {};
 & \node (b-zero) {}; \\
 \node (a-one) {};
 & \node (b-one) {}; \\
 \node (a-two) {};
 & \node (b-two) {}; \\
 \node (a-three) {};
 & \node (b-three) {}; \\[1]
 \node (a-four) {};
 & \node (b-four) {}; \\
 \node (a-five) {};
 & \node (b-five) {}; \\
 \node (a-six) {};
 & \node (b-six) {}; \\
 \node[emptybox] (ae) {};
 & \node[emptybox] (be) {};\\
};
\node [role,draw=none,anchor=south] at (a) {\rname{buyer}};
\node [role,draw=none,anchor=south] at (b) {\rname{seller}};

\draw [dashed] (a.center)--(a-six.center);
\draw [dashed] (b.center)--(b-six.center);

\draw [message] (a-zero.center)--node [mLabelup] {Request} (b-one.center);

\draw [message] (a-two.center)--node [mLabelup] {Payment} (b-three.center);

\draw [message] (b-four.center)--node [mLabelup] {Shipment} (a-five.center);
\end{tikzpicture}
\caption{Payment first.}
\label{fig:payment-first}
\end{subfigure}% <== Mandatory %
%\hfill
\begin{subfigure}[t]{0.33\linewidth}
 \centering
\begin{tikzpicture}
  \matrix () [row sep=10,column sep=70] {
 \node[emptybox] (a) {};
 & \node[emptybox] (b) {};\\
 \node (a-zero) {};
 & \node (b-zero) {}; \\
 \node (a-one) {};
 & \node (b-one) {}; \\
 \node (a-two) {};
 & \node (b-two) {}; \\
 \node (a-three) {};
 & \node (b-three) {}; \\[1]
 \node (a-four) {};
 & \node (b-four) {}; \\
 \node (a-five) {};
 & \node (b-five) {}; \\
 \node (a-six) {};
 & \node (b-six) {}; \\
 \node[emptybox] (ae) {};
 & \node[emptybox] (be) {};\\
};
\node [role,draw=none,anchor=south] at (a) {\rname{buyer}};
\node [role,draw=none,anchor=south] at (b) {\rname{seller}};

\draw [dashed] (a.center)--(a-six.center);
\draw [dashed] (b.center)--(b-six.center);

\draw [message] (a-zero.north)--node [mLabelup] {Request} (b-one.center);

\draw [message] (a-two.north)--node [mLabelup,pos=0.3] {Payment} (b-five.center);

\draw [message] (b-four.center)--node [mLabelup,pos=0.6] {Shipment} (a-six.north);
\end{tikzpicture}
\caption{Concurrent.}
\label{fig:crossing}
\end{subfigure}
\hfill
\caption{Three possible enactments in an ebusiness scenario.}
\label{fig:concurrency}
\end{figure}

We introduce a simple ebusiness scenario in which \rname{buyer} may \mname{Request} some \paraname{item} from a \rname{seller}.
After sending the \mname{Request}, \rname{buyer} may send \mname{Payment} at any time; after receiving \mname{Request}, \rname{seller} may send \mname{Shipment} at any time.
Notice, therefore, that it is possible that \rname{buyer} and \mname{seller} send \mname{Payment} and \mname{Shipment} concurrently.
Indeed, all enactments of Figure~\ref{fig:concurrency} are possible, highlighting the flexibility we desire.

The scenario described above cannot be captured in traditional protocol specification approaches \cite{JAIR-20:Langeval}.
In the information protocols paradigm, it is straightforward to capture it.
Listing~\ref{lst:pay-goods} gives an information protocol for this scenario. In the listing, \inn and \out capture information dependencies.
An agent can send any message whose information dependencies are satisfied by its \emph{local state} (the set of messages sent and received by the agent).
The adornment \inn for a parameter means that the binding for the parameter must exist in the agent's local state; \out means that the binding for the parameter must not exist in the local state, but the act of sending the message generates it.
Thus, enactments of Figure~\ref{fig:concurrency} are all entertained because sending and receiving \mname{Request} satisfy the information dependencies for sending \mname{Payment} and \mname{Shipment}, respectively.

\begin{lstlisting}[basicstyle=\small,caption={An information protocol.}, label={lst:pay-goods},backgroundcolor=\color{listbackgroundcolor}]
Flexible Purchase {
 $\fbf{role}$ B, S
 $\fbf{parameter}$ out ID key, out item, out status, out paid

 B $\mo$ S: Request[out ID, out item]
 S $\mo$ B: Shipment[in ID, in item, out status]
 B $\mo$ S: Payment[in ID, in item, out paid]
}
\end{lstlisting}

Notice that specifying information dependencies as described above means messages can be received in any order.
This operational flexibility truly liberates decision making.

For example, even though \rname{buyer} can emit \mname{Payment} only after emitting \mname{Request}, \rname{seller} can receive \mname{Payment} first.
Further, whenever \rname{seller} receives \mname{Payment}, regardless of whether it has received \mname{Request}, \mname{S} may emit \mname{Shipment} (because its information dependencies would be satisfied).
Further, retransmissions of messages and receptions of duplicate messages are harmless because, information-wise, they are idempotent.
What this means is that information protocols can be flexibly enacted over unordered, lossy communication services such as the Internet.

\subsection{Tango}
Before a protocol may be used to implement agents, we would like to verify that it has certain desirable properties.
Tango is an approach for verifying the safety and liveness of information protocols \cite{IJCAI-21:Tango}.
A protocol is \emph{safe} if no enactment may generate more than one binding for a parameter.
A protocol is \emph{live} if any enactment is able to progress to completion.
Tango is implemented in a command line tool called \val{bspl}, available at \url{https://gitlab.com/masr/bspl/}

To verify a property, we must check that it holds for all possible protocol enactments.
Because information protocols can be flexibly and fully asynchronously enacted (it requires no assumption about message ordering), a protocol may have a large number of enactments, which can make verification inefficient.
An important feature of the Tango approach is that it reduces the set of enactments of protocol to a set of \emph{canonical} enactments and then performs the checking against the set of canonical enactments.
For example, all the enactments in Figure~\ref{fig:concurrency} reduce to one of them.
Thus, instead of checking several enactments, we need to check only one of them.
This reduction leads to vastly improved verification performance compared to earlier verification approaches for information protocols \cite{AAMAS-BSPL-12}.

Listing~\ref{tango:pay-goods} gives the results of executing liveness and safety queries for \pname{Flexible Purchase}, the protocol in Listing~\ref{lst:pay-goods}.
A \emph{maximal} path is one that cannot be extended by further emissions or receptions by an agent.
The third query (\texttt{all\_paths}) makes it clear that there are 12 maximal paths; the tooling reduces them to one for purposes of liveness and safety checking.

\lstdefinestyle{execution}{
  backgroundcolor = \color{white},
  moredelim=**[is][\color{blue}]{@}{@}
}

\begin{lstlisting}[basicstyle=\small\ttfamily,
columns=flexible,breakatwhitespace=false,style=execution, caption={Executing Tango on Flexible Purchase (Listing~\ref{lst:pay-goods}).
Output shown in \color{blue}{blue}.},label={tango:pay-goods}]
>bspl verify liveness Flexible-Purchase.bspl
@{'live': True, 'checked': 7, 'maximal paths': 1, 'elapsed': 0.0010309999343007803}@

>bspl verify safety Flexible-Purchase.bspl
@{'safe': True, 'checked': 7, 'maximal paths': 1, 'elapsed': 0.0009660000214353204}@

>bspl verify all_paths Flexible-Purchase.bspl
@40 paths, longest path: 6, maximal paths: 12, elapsed: 0.005732199992053211
(B!Request, S!Shipment, S?Request, B!Payment, S?Payment, B?Shipment)
(B!Request, S!Shipment, S?Request, B!Payment, B?Shipment, S?Payment)
(B!Request, S!Shipment, S?Request, B?Shipment, B!Payment, S?Payment)
(B!Request, S!Shipment, B?Shipment, B!Payment, S?Payment, S?Request)
(B!Request, S!Shipment, B?Shipment, B!Payment, S?Request, S?Payment)
(B!Request, S!Shipment, B?Shipment, S?Request, B!Payment, S?Payment)
(B!Request, S?Request, B!Payment, S?Payment, S!Shipment, B?Shipment)
(B!Request, S?Request, B!Payment, S!Shipment, B?Shipment, S?Payment)
(B!Request, S?Request, B!Payment, S!Shipment, S?Payment, B?Shipment)
(B!Request, S?Request, S!Shipment, B!Payment, S?Payment, B?Shipment)
(B!Request, S?Request, S!Shipment, B!Payment, B?Shipment, S?Payment)
(B!Request, S?Request, S!Shipment, B?Shipment, B!Payment, S?Payment)@
\end{lstlisting}

Listing~\ref{lst:buggy-pay-goods} gives a buggy variant of \pname{Flexible Purchase}.
It is not live because \paraname{status}, which must be bound for the completion, cannot be bound in any enactment.
It is unsafe because both \rname{B} and \rname{S} can send \mname{Payment} and \mname{Shipment} concurrently, thus concurrently binding \paraname{paid}.

\begin{lstlisting}[basicstyle=\small,caption={An unsafe and nonlive version of the protocol in Listing~\ref{lst:pay-goods}.}, label={lst:buggy-pay-goods},backgroundcolor=\color{listbackgroundcolor}]
Buggy Flexible Purchase {
  roles B, S
  parameters out ID key, out item, out status, out paid 
  
  B -> S: Request[out ID key, out item]
  S -> B: Shipment[in ID key, in item, out paid]
  B -> S: Payment[in ID key, in item, out paid]
}
\end{lstlisting}

Listing~\ref{tango:buggy-pay-goods} shows the outcomes of liveness and safety queries for this protocol, giving counterexamples for both.
For liveness, it gives an enactment that is deadlocked.
For safety, it reports the parameter (\paraname{paid}) as being multiply bound.

\begin{lstlisting}[basicstyle=\small\ttfamily,
columns=flexible,breakatwhitespace=false,style=execution, caption={Executing Tango on \pname{Buggy Flexible Purchase}.},label={tango:buggy-pay-goods}]
>bspl verify liveness Buggy-Flexible-Purchase.bspl
@{'live': False, 'reason': 'Found path that does not extend to completion', 'path': (B!Request, B!Payment, S?Payment, S?Request), 'checked': 5, 'maximal paths': 1, 'elapsed': 0.001316600013524294}@

>bspl verify safety Buggy-Flexible-Purchase.bspl
@{'safe': False, 'reason': 'Found parameter with multiple sources in a path', 'path': (B!Request, B!Payment, S?Request, S!Shipment), 'parameter': 'paid', 'checked': 7, 'maximal paths': 1, 'elapsed': 0.0017665999475866556}@
\end{lstlisting}

\subsection{Kiko}
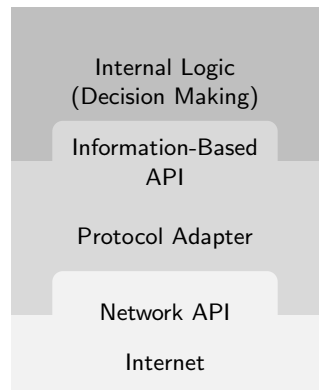
\begin{figure}[htb!]
 \centering
\begin{tikzpicture}[>=stealth]

\tikzstyle{box}=[draw=none,rounded corners,align=center,font=\sffamily,fill=blue!80!gray!20,rectangle,anchor=center,minimum height=8ex,minimum width=22ex,inner sep=2]
\tikzstyle{napibox}=[draw=none,rounded corners,align=center,font=\sffamily,fill=gray!10,rectangle,anchor=center,minimum height=8ex,minimum width=22ex,inner sep=2]

\tikzstyle{abox}=[draw=none,sharp corners,align=center,font=\sffamily,fill=blue!80!gray!30,rectangle,anchor=center,minimum height=15ex,minimum width=30ex,inner sep=2]
\tikzstyle{bbox}=[draw=none,sharp corners,align=center,font=\sffamily,fill=blue!80!gray!20,rectangle,anchor=center,minimum height=15ex,minimum width=30ex,inner sep=2]
\tikzstyle{cbox}=[sharp corners,align=center,font=\sffamily,fill=blue!80!gray!10,rectangle,anchor=center,minimum height=8ex,minimum width=30ex,inner sep=2]
\tikzstyle{ebox}=[draw=none,sharp corners,align=center,font=\sffamily,fill=blue!20!gray!40,rectangle,anchor=center,minimum height=15ex,minimum width=30ex,inner sep=2]
\tikzstyle{edge_label_org}=[draw=none, pos=0, fill=white,inner sep=2,font=\sffamily]

\matrix (swiml) [draw=none,fill=none,rounded corners,row sep=0,
  column sep=-\pgflinewidth] {
   %\node[ebox,fill=none] {}; \\
   %
   \node[abox,fill=gray!50] (dml) {Internal Logic\\ (Decision Making)}; \\
   \node[bbox,fill=gray!30] (mcl) {Protocol Adapter}; \\
   \node[cbox][fill=gray!10](cic) {\\Internet};\\
};
%\draw (mcl)--node [edge_label_org,midway, align=center] (meaning) {\lang{}\\Spec.} (mcr);
\node[box,fill=gray!30](apil) at ($(dml.south)+(0,-0)$) {Information-Based \\API};

\node[napibox](capil) at ($(dml.south)+(0,-2)$) {Network API};
\end{tikzpicture}
\caption{Agent architecture in the Kiko programming model.}
\label{fig:kiko}
\end{figure}

Kiko is a protocol-based programming model for agents.
Specifically, given an information protocol, it enables implementing agents that play roles in the protocol.
To make agent development easy, Kiko includes a middleware that exposes an event-driven, information-based interface that may be used to implement an agent's internal reasoning.

\begin{figure}[htb!]
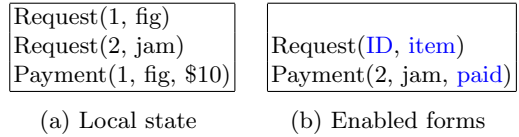

\centering
\subfloat[Local state]{
\begin{tabular}{|l|}\hline
Request(1, fig)\\
Request(2, jam)\\
Payment(1, fig, \$10)\\\hline
\end{tabular}}
\quad
\subfloat[Enabled forms]{
\begin{tabular}{|l|}\hline
\\
Request(\textcolor{blue}{ID}, \textcolor{blue}{item})\\
Payment(2, jam, \textcolor{blue}{paid})\\\hline
\end{tabular}}
\caption{A possible local state for a \rname{buyer} agent and the enabled forms in that state.}
\label{fig:forms}
\end{figure}

As Figure~\ref{fig:kiko} shows, each Kiko agent has an information protocol adapter that sits between the network and the agent's decision making, that is, its internal logic.
An agent's Kiko adapter maintains its local state.
Based on the local state and the protocol specification, it keeps track of information-enabled \emph{forms}.
The forms are necessarily partial message instances that would be legal to send if completed.
Specifically, a form's \inn parameters are bound (from the local state), and the \out parameters are unbound (because they don't exist in the local state).
(Information protocols may also feature the \nil adornment, which we omit from this discussion for simplicity.) 
Figure~\ref{fig:forms} gives a possible local state for a \rname{buyer} agent and the forms available to it in that state.

To create a Kiko agent, a developer writes a set of \emph{decision makers}.
A decision maker is an event-triggered piece of code that gets the set of enabled forms and completes some subset via some logic.
The completed forms are emitted by the adapter as messages and added to the local state.

\begin{lstlisting}[caption={A \rname{buyer} agent's decision maker that sends \mname{Payment} only in those enactments (as identified by \paraname{ID}) in which \mname{Shipment} has been received.},label={list:decision-maker}]
@adapter.schedule_decision(00 17 * * *)
def payment(enabled, state):
  payments = enabled.messages(Payment)
  for p in payments: 
     if(next(state.messages(Shipment, system=p.system, ID=p["ID"])))
        p.bind(paid="10")
\end{lstlisting}

Listing~\ref{list:decision-maker} shows a decision maker for a \rname{buyer} agent.
Its logic is to complete those \mname{Payment} forms for which \mname{Shipment} has been received.
The completed \mname{Payment} forms are sent by the adapter as messages.
The decision maker is triggered at 1700 hours every day.
In other words, it processes enabled \mname{Payment}s in a batch every day.

Listing~\ref{list:adapter} informally describes the logic of the adapter.
In essence, the adapter runs a loop in which it either responds to a trigger by invoking the corresponding decision maker or receives a message (if one is available).

\begin{lstlisting}[caption={Adapter logic.},label={list:adapter}]
//$t_i:d_i$ represents a decision maker $d_i$ with trigger $t_i$
//$c_i$ represents a channel on which messages from another agent may be received
while() 
    r = listen($t_1$,$\ldots$,$t_m$, $c_1$,$\ldots$,$c_n$)
    if (r is a $t_i$)  
        attempts = invoke($d_i$, forms)
        if (check(attempts))
            updateLocalState(attempts)
            emit(attempts)
    else
        m = receive(r)
        if(check(m))
            updateLocalState(m)
\end{lstlisting}

\emph{Attempts} refer to completed forms. This terminology reflects the fact that completed forms may be mutually inconsistent and therefore must be checked by the adapter before emission.
Notice how the adapter abstracts away the actual emission and reception of messages from the agent developer.
Unlike alternative approaches, a Kiko developer never needs to \emph{selectively receive} messages---that is, specify which message to receive next and block until it arrives.
As in the actor model \cite{Hewitt77}, messages are received and added to the local state as they are made available by the network.
A feature of Kiko is that it needs nothing more than UDP (part of the Internet Protocol suite) for transport.
UDP, notably, supports only best-effort (unreliable, unordered) message delivery.

The Kiko adapter, documentation, and sample code are available at:

\href{https://gitlab.com/masr/bspl/-/tree/kiko/}{https://gitlab.com/masr/bspl/-/tree/kiko/}

\subsection{Mandrake}

Mandrake is another protocol-based programming model.
Kiko's enablement-based programming is more general and convenient to use than Mandrake's \emph{reactive} programming model (an agent reacts to the reception of a message by sending a message).
Where Mandrake goes beyond Kiko is in the handling of delayed (potentially lost) messages, a kind of fault that may arise when using network transports such as UDP.
Specifically, Mandrake enables writing agent-level policies for dealing with such faults.

Mandrake is inspired by the end-to-end principle \cite{Saltzer+84}.
Although it is today customary to assume a reliable transport layer such as TCP (also part of the Internet Protocol suite), the end-to-end principle advises that lower-layer reliability guarantees are inadequate for building reliable multiagent systems.
Instead, the agents must encode the necessary reliability mechanisms.

A simple example highlights the inadequacy of lower-level reliability guarantees.
Suppose the \rname{seller} agent has sent \mname{Shipment} but has not received \mname{Payment} even after waiting a considerable amount of time.
There are two reasons why the \rname{seller} has not received \mname{Payment}.
Either \rname{buyer} sent the \mname{Payment} but it was delayed in transit, or \rname{buyer} never sent \mname{Payment}.
These reasons are indistinguishable to the \rname{seller}.

Whereas TCP can help address network problems via retransmission, it is of no help if an agent never sends a message.
Therefore, the \rname{seller} must implement some fault handling logic.
For example, as is often the case in real life, it may remind the \rname{buyer} about its \mname{Shipment}.
Listing~\ref{list:reminder} shows such a retransmission policy for a \rname{seller} agent.
The policy specifically is to send a daily reminder of the \mname{Shipment} to the \rname{buyer} until it receives the \mname{Payment} or it has sent five reminders.

\begin{lstlisting}[caption={A \rname{seller}'s reminder policy.},label={list:reminder}]
- action: remind Buyer of Shipment until Payment
  when: 0 0 $*$ $*$ $*$ // daily
  max tries: 5
\end{lstlisting}

Notice that the reminders amount to application-level retransmission of messages.
As explained above, they are necessary, but once we have them at the application level, lower-level reliability mechanisms, such as those provided by TCP, are obviated, thus leading to potentially improved performance.
Mandrake is available at \url{https://gitlab.com/masr/bspl/-/tree/mandrake/}

\subsection{Reflections}

The autonomy of entities motivates protocols, which are thus inherent to multiagent systems.
There are, of course, multiagent systems that are implemented without \emph{specifying} a protocol.
In such cases, the agent programmer implements the protocol in low-level agent code.
For anything but the most trivial protocols, such an exercise is bound to be complex and error-prone.
By enabling the specification of protocols and implementing protocol-based agents, our tooling addresses this gap.

We started with a simple protocol \pname{Flexible Purchase}; verified it for safety and liveness using Tango, and showed how one might implement agents using Kiko and Mandrake.
Keeping in mind the alternative approaches described in Section~\ref{sec:literature}, the following points are noteworthy about our approach.

\begin{itemize}

\item Our approach is formal, and our tooling enables implementing a verified protocol.
That is, the model you verify is the model you implement.

\item \pname{Flexible Purchase}, as simple as it is, is not even expressible as a well-formed protocol in alternative protocol specification approaches \cite{JAIR-20:Langeval}.

\item In general, protocol specifications can be much more complex than our examples.
The number of possible enactments of a protocol can, in the worst case, grow exponentially with the size of the protocol.
Kiko's enablement-based approach abstracts away much of this complexity.

\item Our approach does not require message ordering guarantees from the communication infrastructure.
Both the actor model \cite{Agha86,Hewitt77} and the end-to-end principle \cite{Saltzer+84} argue against such guarantees.
Mandrake enables agent programmers to encode policies for handling potential communication failures in a straightforward manner.

\item Our approach leads to decentralized multiagent systems---agents communicate with each other via asynchronous messaging.
A ``centralized'' multiagent system would be a special case in which the protocol is such that there is one role such that all communications are between it and the other roles.

\item High-level meaning is important.
Although not the focus of this contribution, Table~\ref{tab:tools} gives pointers to tools that show how protocols are a substrate that supports meaning.
Both are necessary: whereas protocols define the legal communicative moves with respect to causality and integrity, notions such as commitments capture their meaning.

\end{itemize}

\section{Summary of Tools for Interaction-Oriented Programming}

This paper introduces publicly available software that embodies the principles of Interaction-Oriented Programming  (IOP) and should help facilitate the adoption of multiagent systems by developers.
The software is centered on the idea of information protocols, a novel approach for modeling flexible multiagent systems.
The software enables verifying protocols and implementing protocol-based agents.
Table~\ref{tab:tools} summarizes the various tools for engineering multiagent systems based on models of interaction that we have developed and which are still active.
We have explained Tango, Kiko, and Mandrake above.
We briefly explain the others below.

\begin{table}[htb]
\centering
\caption{Tools for engineering multiagent systems based on interaction: information protocols and commitments.
Internals refers to the language the tool is constructed using and Usage refers to the language it outputs or analyzes.}
\label{tab:tools}
% \begin{tabular}{l p{4.0cm} l l l}\toprule
\begin{tabular}{l l l l l}\toprule
&& \multicolumn{2}{c}{\fbf{Language}}\\\cmidrule{3-4}
\fbf{Tool} & \fbf{Main purpose} & Internals & Usage & \fbf{Status}\\\midrule

% \multicolumn{5}{c}{Currently active tools}\\\cmidrule(lr){1-5}

Tango & Verifying protocols & Python & BSPL & Stable \\

% Mambo & Verifying protocols with custom queries & Python & BSPL & Alpha \\

Kiko & Enablement-based imperative agent programming & Python & Python & Stable\\

Mandrake & Reactive, fault-tolerant agent programming & Python & Python & Stable\\

Orpheus & Enablement-based cognitive agent programming & Jason & Jason & Beta \\

Cupid & Compile commitments into database queries & Cupid & SQL & Beta \\

Azorus & Enablement-based cognitive agent programming with commitments & Jason & Jason & Alpha \\
% \midrule

% \multicolumn{5}{c}{Previous tools, no longer active}\\\cmidrule(lr){1-5}

% BSPL & Verifying protocols & Java & BSPL & Obsolete \\

% Atomic & Verifying protocols for atomicity & Java & BSPL & Obsolete \\

% Clouseau & Synthesizing protocols from commitments & Java & Cupid & Alpha

\bottomrule
\end{tabular}
\end{table}

Orpheus \cite{AAAI-25:Orpheus} is a programming model inspired by Kiko but geared toward implementing cognitive agents in Jason.
Orpheus provides a tool that generates much of the general-purpose reasoning needed by an agent that is based on the given protocol (and the BSPL semantics).
A programmer needs to provide only the business logic comprising the agent's goals (presumably to support the agent's principal's requirements) and the mapping of these goals to the messages the agent needs to send.

Communication meaning is a defining theme in multiagent systems.
A superior alternative to the FIPA ACL \cite{FIPA-02:ACL} and KQML \cite{KQML92} is specifying social meaning \cite{Ijcai-99-ACL} in terms of commitments and kinds of norms.
Cupid is a language for specifying the meaning of communicative acts in terms of commitments \cite{AAAI-15:Cupid}.
Cupid is accompanied by a compiler that translates commitment specifications into SQL queries over a database of communicative acts.

Azorus \cite{AAMAS-25:Azorus} extends Orpheus by incorporating support for commitment-based reasoning in Jason.
Specifically, Azorus includes a Cupid compiler targeted to Jason.
Jason programmers can now write agents that query states of commitments and take protocol-based actions accordingly.
Interestingly, though commitment protocols \cite{Aamas-99,Atal-01} precede information protocols by over a decade, the connection between commitments and information protocols for the purpose of programming multiagent systems was not fleshed out until recently.
Azorus begins to fill that gap.

\section{Artificial Intelligence Context}

Our current techniques fall squarely in the so-called Good Old-Fashioned AI (GOFAI) camp in that we seek to model and verify multiagent systems on the basis of requirements and then implement them using middleware-supported programming abstractions.

We seek representations that are intuitive to both stakeholders and programmers.
Our whole enterprise is, in fact, motivated by the need for explicitly modeling the high-level meaning of interactions via notions such as commitments and other norms.
It is worth keeping in mind that interoperability is an important goal of engineering multiagent systems, and because interoperability between agents depends on them having a common understanding of such meaning, there is no alternative to specifying it explicitly.
In this regard, ours is no different from any existing multiagent systems engineering approach.
Our approach has room for exploiting semantic descriptions of objects \cite{Lemee:signifiers:2024} and planning-based agents \cite{AAAI-HTN-13,AAMAS-HTN-13}.

The Theory of Mind (ToM) is the notion that agents model other agents as having minds \cite{Carruthers+Smith-96:ToM-intro,Premack+Woodruff-78:ToM-chimps}.
ToM justifies the use of folk psychological concepts (e.g., beliefs and goals) in AI, e.g., \cite{Dennett87,Newell82,Self94:book}.

Although the current paper deemphasizes representing agents in terms of BDI, our other work (discussed above) addresses this theme.
There is also an opportunity to combine our approach with learning agents \cite{Sutton+Barto-18:RL,IJCAI-22:XSIGA}.
For example, an agent may learn to act in ways that represent violations of the specified norms but lead to overall societal benefit.
We are agnostic on the representation of an agent's decision making except where it interfaces with communications.
Those communications must be made based on protocols (for interoperability) and in the light of their specified meaning (ignoring which could result in unexpected sanctions and missed opportunities).

Large Language Models (LLMs) and other generative AI have upended many intellectual fields.
LLMs perform comparably to humans in tasks such as question answering and code generation \cite{Mozannar+24:coding-AI}, though with well-known limitations in reasoning \cite{Verma-24:brittleness-LLMs}.

The \emph{agentic} methods form ``agents'' by combining generative AI with information and abilities to sense and act, e.g., via web services \cite{Salesforce-24:xLAM}.
Current frameworks emphasize workflows to execute compositions of agents specified as task graphs \cite{LangGraph-24,Microsoft-24:AutoGen-Studio}.
Formulating agent coordination in task graphs sounds attractive, but its shortcomings have been known for decades \cite{ACTA94,Expert94,Vldb93}.
Choreographies \cite{Peltz-03} go beyond workflows in considering multiple loci of action, and they too are limited \cite{AAMAS-BSPL-11,AAMAS-BSPL-12}.

A major shortcoming of workflows and choreographies is their rigidity, which limits agent autonomy and responsiveness to exceptions \cite{ICSOC-24:choreographies}.
Thus, their adoption in modern agentic frameworks is particularly unfortunate---the capabilities of agents built with modern AI techniques would be stymied by poor coordination.
Thus, a challenge is to develop agents who can reason about and interact based on protocols.
LLMs have shown capabilities comparable to humans on some ToM tasks, mostly focused on beliefs \cite{Kosinski-24:evaluating-ToM-LLM,Strachan-24:ToM-LLM-humans}, whereas humans (even children) understand desires more readily than beliefs \cite{Harris-96:ToM}.
Enhancements to LLMs for collaboration via ToM could help go beyond today's trivial agentic models.

\section{Blue Skies All Around!}

A vision that inspired the field of multiagent systems was that agents would better capture autonomy in distributed systems.  A key enabling idea in this direction was that of interaction meaning: If agents could understand the meaning of their interactions, then they could make decisions intelligently. Early approaches emphasize mentalist semantics, e.g., as in KQML and FIPA ACL.  Despite their many and well-documented shortcomings (as discussed in Section~\ref{sec:literature}), their usage continues within popular multiagent programming frameworks.  The work highlighted in this paper highlights a practical basis for intelligent, decentralized decision making without the shortcomings of the mentalist approaches. A direction for the entire engineering multiagent systems (EMAS) community is to reorient their abstractions and tooling to incorporate protocols and norms.  Additionally, we need standards for developing multiagent systems that are based on these ideas. We think such standards would galvanize the EMAS community and make its work attractive to practitioners. 

Interaction-Oriented Programming is very much an ongoing effort, and we expect to augment and improve the software suite over the coming years. We highlight some specific directions here.  An additional Blue Sky discussion is in \cite{chopra:blue-sky:2023}.

\emph{Higher-Level Protocol Languages.} Although BSPL is foundational and higher-level than alternative protocol languages, it captures individual messages and may be thought of as an assembly language for operational protocols.
In particular, it does not distinguish information that is essential to message meaning from information that is purely coordinative.
For example, in the context of an alternative purchase protocol, imagine that a \rname{buyer} may send either \mname{Accept} or \mname{Reject} in response to an \mname{Offer}.  
Parameters such as \paraname{ID}, \paraname{item}, \paraname{price} are related to the meaning of \mname{Accept} and \mname{Reject} and thus would feature in both.
To make \mname{Accept} and \mname{Reject} mutually exclusive in BSPL, we would need another parameter which would be \out in both.
This parameter is purely coordinative; it has nothing to do with the meanings of \mname{Accept} and \mname{Reject}.
The coordination requirements could, of course, be more complex than mutual exclusion, and a protocol designer would have to map those requirements to the elementary notions in BSPL.
This motivates protocol languages that are higher-level than BSPL in that they are focused on meaning and enable generating the necessary coordination.
Langshaw \cite{IJCAI-24:Langshaw} is a start in that direction. Langshaw protocols are synchronous and specify agent actions and \emph{sayso}s (the capability of generating it) over information. BSPL protocols are compiled out from Langshaw protocols. 

\emph{Fault Tolerance and scalability.} These are both of practical importance but are not particularly well-addressed in the EMAS community. 
Mandrake requires the specification of fault tolerance policies.
Ideally, we should exploit the meaning specifications to automate fault tolerance.
For example, commitment specifications could tell an agent when and which communications are important and, therefore, worth retransmitting.

Modern architectural paradigms, such as the cloud, are motivated by scalability.
Though there have been efforts linking information protocols and cloud computing paradigms such as serverless computing \cite{ICWS-21:Deserv} and microservices \cite{EMAS-23:Dapr}, research on engineering multiagent systems, including IOP, lacks a story for how to realize highly scalable multiagent systems.
The actor model is known for its scalability; developing a synthesis of protocols and actors would be a valuable direction.

\emph{Types.} A type theory for protocols would further enhance the programming mode by catching errors at compile time and guiding the implementation of agents.  For example, we could capture at compile time the error of an agent attempting to send both \mname{Accept} and \mname{Reject} in an enactment. The notion of dependent types (e.g., as in Idris \cite{Idris-25}) could be interesting for this purpose.

\emph{Methodologies.} Developing methodologies for specifying meaning, protocols, and implementing agents will be important.
We need to understand how requirements map to the specification of meaning in terms of norms, e.g., in the spirit of Protos \cite{RE-14:Protos}.
We need to extend Tango to support the verification of a broader class of properties, e.g., encoding the stakeholder requirements on protocols.
We need to understand how to map agent requirements into agent code in programming models such as Kiko.
Kiko guarantees an agent's compliance with the protocol--- statically, in the case of sequential agents.
We may also want to verify that the agent's internal logic meets requirements via formal methods and testing.

\section{Reproducibility}
The entire codebase referenced in this paper, as well as other related tools, are available online at
\url{https://gitlab.com/masr}.
Software developed in \cite{EMAS-23:Dapr} is available at \url{https://gitlab.com/masr/information-protocols-dapr-emas-2022}.
Software developed in \cite{ICWS-21:Deserv} is available at \url{https://gitlab.com/masr/deserv}.

\section*{Acknowledgments}
Thanks to the NSF (grant IIS-1908374) and EPSRC (grant EP/N027965/1) for support.

\bibliographystyle{splncs04}
%\bibliography{Amit,Munindar}

\DeclareRobustCommand{\nUmErAL}[1]{#1}\DeclareRobustCommand{\nAmE}[3]{#3}\DeclareRobustCommand{\nUmErAL}[1]{#1}\DeclareRobustCommand{\nAmE}[3]{#3}

\end{document}